\documentclass[galley,grl]{agu2001}
\usepackage{graphicx}

\authorrunninghead{PETITDEMANGE ET AL.}

\titlerunninghead{MRI in the Earth's outer core}

\authoraddr{Ludovic Petitdemange, Emmanuel Dormy, Steven Balbus,
LRA, D\'epartement de Physique, 
Ecole Normale Sup\'erieure,
24, rue Lhomond,
75231 Paris Cedex 05, France
(dormy@phys.ens.fr)}

\begin{document}

%
%
%
%
%

%
%

\newcommand\bb[1] {   \mbox{\boldmath{$#1$}}  }
\newcommand\beq{ \begin{equation} }
\newcommand\eeq{ \end{equation} }
\newcommand\kva{ \bb{k\cdot V_A}  }
\newcommand\dd{\partial}
\newcommand\vv{\bb{V}}
\newcommand\vvv{\bb{v'}}
\newcommand\B{\bb{B}}
\newcommand\bcdot{\bb{\cdot}}
\newcommand\btimes{\bb{\times}}
\newcommand\del{\bb{\nabla}}
\newcommand\bfnabla{\bb{\nabla}}

%
%

\title{Magnetostrophic MRI\\ in the Earth's Outer Core}

%
%

\authors{Ludovic Petitdemange, \altaffilmark{1,2}
Emmanuel Dormy, \altaffilmark{1,3}
and Steven A. Balbus \altaffilmark{1,2}}

\altaffiltext{1}
{MAG (ENS/IPGP)\break
LRA, D\'epartement de Physique\break
Ecole Normale Sup\'erieure\break
24, rue Lhomond\break
75231 Paris Cedex 05, France.}

\altaffiltext{2}{LERMA, Observatoire de Paris, CNRS/UMR8112.}

\altaffiltext{3}{Institut de Physique du Globe de Paris, CNRS/UMR7154.}

%
%


\begin{abstract}
We show that a simple, modified version of the Magnetorotational Instability
(MRI) can develop in the outer
liquid core of the Earth, in the presence of a background shear. 
It requires either thermal wind, or a primary
instability, such as convection, to drive a weak differential rotation within 
the core.
The force balance in the Earth's core is very unlike
classical astrophysical applications of the MRI (such as gaseous disks
around stars). 
Here, the weak differential rotation in the Earth core yields an 
instability by its        
constructive interaction with the planet's much larger rotation
rate. The resulting
destabilising mechanism is just strong enough to counteract stabilizing
resistive effects, and produce growth on geophysically interesting
timescales.
We give a simple physical explanation of the instability, 
and show that it relies on 
a force balance appropriate to the Earth's core, known as            
magnetostrophic balance. 

\end{abstract}

%
%

%

\begin{article}

%
%

\section{Introduction}
The Magnetorotational Instability (MRI) is important for
differentially rotating astrophysical 
objects such as gaseous disks, because it forces
a breakdown of laminar flow into turbulence,
producing the enhanced dissipation and transfer of angular 
momentum necessary for accretion of matter onto the central
massive object (Balbus and Hawley, 1991). 
Disk investigators began with purely hydrodynamical candidate mechanisms,
but now the center of interest is squarely upon
magnetohydrodynamics.   Interest in the Earth's core, by contrast,
has been magnetic almost from the start.  
The principal problem, of course, has been to understand how fluid motions in the
core generate a magnetic field.  
This is often approached via {\em kinematic} processes in which
a hydrodynamically turbulent fluid exponentially amplifies a 
very weak seed magnetic field. 
Here, we argue of a version of the MRI, dynamically
coupling both the velocity and the magnetic field,
can also provide a mechanism for linear instability in the Earth's
core. While this mechanism is not meant to serve as the primary dynamo 
process, it could be an important source of secondary instabilities and 
magnetic secular variation.

We are hardly the first authors to study the effects of differential
rotation on the dynamical stability of the geodynamo (see e.g. Acheson, 1983;
Ogden and Fearn, 1995; Fearn et al. 1997).
But in previous calculations
the emphasis has been upon purely azimuthal fields,
nonaxisymmetric disturbances, and magnetic instabilities.
In this work, the dynamical focus is much different.  
Here the magnetic coupling
is to the poloidal field components,
axisymmetric disturbances
are front and center,
and the instability, while relying on the presence
of a magnetic field, has its seat of free energy entirely in differential
rotation.  The MRI is a somewhat novel concept in this context,
and is worthy of study in isolation.  Fortunately, it
can be understood in very direct and simple physical terms. 

The Earth's core is, by comparison to accretion disks, a relatively
small object, in which resistive effects are on an equal footing
with dynamical processes.  The rotation properties of the Earth also
significantly differ from those of an accretion disk. To leading 
order, they correspond to solid body rotation, with only a 
weak differential rotation.  At first sight, it is far from
obvious that the Earth's core is a venue for the MRI.  

The purpose of this {\it Letter} is to show that, despite its weak
differential rotation and significant resistive effects,
the Earth's core can in fact host the MRI.  We derive a 
WKB dispersion relation relevant to this asymptotic regime (i.e. 
magnetostrophic balance), and demonstrate 
excellent agreement between this local description and global
numerical simulations in spherical geometry.  The final section
speculates on potential applications of our results.

\section{Modelling}  
The physical 
parameter regime relevant to the Earth's outer core dynamics is relatively 
well constrained. In order to investigate the MRI in a conducting fluid, it is
also necessary to provide a reasonable approximation to the basis flow profile
over which the instability may develop. 
Apart from very restricted components of the velocity field,
imaging of the core flow is limited to surface flows
(e.g. Bloxham \& Jackson 1991;
Amit \& Olson, 2004; Eymin \& Hulot, 2005).   Moreover, while secular magnetic field variations
allow for an order-of-magnitude estimate for the zonal flow
(Pais and Hulot 2000), it is not possible at present to
achieve a direct, detailed or reconstructive imaging
of the global differential rotation of the Earth's core.
Asymptotic studies early identified the essential role of the zonal shear
in the geodynamo (Taylor, 1963), and this has remained central to
our understanding of outer core dynamics. Numerical models of
convection in rotating spheres (e.g. Christensen, 2002) or in convectively 
driven dynamos (e.g. Aubert, 2004), also observed the presence of a 
strong zonal shear.  Besides surface flow reconstructions,
the only (indirect) observational
evidence for such shear is
inferred from seismological data that have been interpreted 
as rotation of the solid inner core at a rate of about
$0.15^{\rm o}$ per year relative to the mantle (Vidale, Dodge \& Earle, 2000;
see also Dumberry 2007). 
But is this change in the angular velocity profile spread across the 
whole core radius or, say, localised in a narrow shear layer?  There are
no observational constraints on the way the
corresponding jump in angular velocity is actually accommodated
by the flow.

\section{Magnetostrophic-MRI local description}

Consider the stability of a differentially rotating fluid, with
angular velocity $\Omega$ a function of $s$ and $z$ in a standard
cylindrical coordinate system $(s, \phi, z)$.  The density $\rho$,
pressure $P$,
magnetic field $\bb{B}$ are functions of $s$ and $z$
in the equilibrium state, and of course
depend on time $t$ as well when the
equilibrium is perturbed.  The fluid is characterized by
a kinematic viscosity $\nu$ and resistivity $\eta$.
We work in the WKB limit, assuming that all
linearized perturbations have the space-time dependence
\beq
Q \propto \exp  (ik_s s + ik_z z +\sigma t)
\eeq
where $Q$ is an infinitesimal (Eulerian) disturbance in a fluid quantity.  
Our starting point is the general dispersion relation of a constant
density fluid adapted from
Menou, Balbus, \& Spruit (2004):
\beq
{k^2\over k_z^2} {\widetilde\sigma}_{\eta\nu}^4
-{\widetilde\sigma}_{\eta\eta}^2
\left[
{1\over s^3}\, {\cal D} (s^4\Omega^2) \right] - 4 \Omega^2 (\kva)^2=      
0\,,
\eeq
where 
\beq
{\cal D} = \left( {k_s\over k_z}{\dd \over \dd z} -{\dd\ \over \dd s}
\right)\, ,
\eeq
and
\beq
{\widetilde\sigma}_{\eta\nu}^2 = \left[
(\sigma +\eta k^2)(\sigma +\nu k^2) +(\kva)^2 \right]\, ,
\eeq
\beq
{\widetilde\sigma}_{\eta\eta}^2 = \left[
(\sigma +\eta k^2)^2 +(\kva)^2 \right]\, .
\eeq
The Alfv\'en velocity $V_A$ is defined as
\beq
\bb{V_A}= {\bb{B}\over \sqrt{\mu\rho}}
\eeq
where $\mu$ is the magnetic permitivity.  Finally, we note that
$\Omega$ is the full angular velocity, i.e. if $\Omega_0$
denotes the Earth rotation rate, and $\bb{V}$ the velocity in the
rotating frame, then $\Omega=\Omega_0+V_\phi /s$~.

To see whether the unstable MRI modes can be present in a very simple
model, we consider the case of an initial poloidal magnetic field
which is locally vertical and $\Omega$ is a function only of $s$. 
(We stress here that we use a vertical magnetic field for the
sake of simplicity of the presentation, but that the overall conclusion can
be easily extended to the case of more general applied fields.)
Then
\beq
{\cal D} (s^4\Omega^2)= -\left(4\Omega^2 + s{d\Omega^2\over d s}\right)
\equiv -\kappa^2 \,.
\eeq
For the current application, the shear gradient is very small compared
with $4\Omega^2$, and $\kappa$, which is known as the ``epicyclic frequency''
in the astrophysical literature, is very nearly $2\Omega$.
With these
assumptions, the dispersion relation becomes
\beq
{k^2\over k_z^2}\left[ (\sigma+\eta k^2)(\sigma+\nu k^2) +(\kva)^2\right]^2
+4\Omega^2(\sigma+\eta k^2)^2 +(\kva)^2 s{d\Omega^2\over d s}=0 \, .
\label{complete_rel_disp}
\eeq
This is the form that we have used (see below)
for comparison with numerical simulations.  

In geodynamo applications, the viscosity is very small compared with the
resistivity (the ratio is $10^{-5}-10^{-6}$) and it may be dropped.
The instability of interest arises from the final term of this dispersion
relation, if the angular velocity decreases outwards.  
This is a small
term in the sense that $\Omega$ is regarded as large,
and we expect that the growth rate $\sigma$
will itself be very small compared with $\Omega$.  
We may therefore drop all $\sigma$ terms from our dispersion relation,
{\em except} for those which are multiplied by $\Omega^2$.  (This
procedure may be formalized by normalizing $\sigma$ with respect
to $|s\,d\Omega/d s|$ and then treating 
$$
\varepsilon \equiv  {s\over \Omega}\,{d\Omega\over d s} \sim 10^{-6}
\, ,
$$
as a vanishingly small expansion parameter.)  We obtain
\beq\label{disp}
{k^2\over k_z^2}(\kva)^4 +4\Omega^2(\sigma+\eta k^2)^2 
+(\kva)^2 s\, {d\Omega^2\over d s} = 0.
\eeq

This relatively simple dispersion relation has a correspondingly simple
physical interpretation.  Consider
magnetostrophic balance with an axial 
wave number ($\bb{k}=k\bb{e_z}$),
\beq
2\bb{\Omega\times V} = -\del \Pi +\frac{1}{\mu \rho}(\B\bcdot\del)\B\, ,
\eeq
where $\Pi$ includes the material and magnetic pressures as well as the
centrifugal potential,
along with the full induction equation
\beq
\left( {\dd\ \over \dd t} + \vv\bcdot\del\right)\B =(\B\bcdot\del)\vv
+\eta\nabla^2\B \, .
\eeq
The equilibrium profile is to leading order
$V_\phi=s[\Omega_0 + (s-s_0)\, d\Omega/ds]$,
and $\B=B\bb{e_z}$.
Consider linear velocity perturbations $ v_s$ and $ v_\phi$,
and magnetic field perturbations $ b_s$
and $ b_\phi$ of the form $\exp(\sigma t+ikz)$.  
In magnetostrophic
balance, the radial and azimuthal linearized equations are respectively
\beq\label{ms1}
-2\Omega  v_\phi = {ikB_0\over \mu_0\rho} \,  b_s\,,
\eeq
\beq\label{ms2}
{2\Omega}  v_s ={ikB_0\over \mu_0\rho} \,  b_\phi\,.
\eeq
The same components of the induction equation are
\beq\label{ms3}
(\sigma +\eta k^2)\,b_s = ikB_0 v_s\,,
\eeq
\beq\label{ms4}
(\sigma +\eta k^2)\, b_\phi = ikB_0\, v_\phi + s \, {d\Omega\over d s}\, b_s\,.
\eeq
Combining (\ref{ms3}) and (\ref{ms2}) leads to
\beq\label{fftn}
(\sigma +\eta k^2)\, b_s= - {1\over 2\Omega} (kV_A)^2
b_\phi\, ,
\eeq
whereas (\ref{ms4}) and (\ref{ms1}) imply
\beq\label{sxtn}
(\sigma +\eta k^2)b_\phi = \left[
{(kV_A)^2 \over 2\Omega} + s\, {d\Omega\over d s} \right] b_s
\, .
\eeq
Notice the absence of pressure-like perturbations for axial wavenumbers.
The last two equations combine to give
\beq
(kV_A)^4 +4\Omega^2(\sigma+\eta k^2)^2 +(kV_A)^2 s {d\Omega^2\over ds}
=0,
\eeq
which is just equation (\ref{disp}) for axial wavenumbers.

By restricting the perturbations exclusively
to the magnetic field components, equations
(\ref{fftn}) and (\ref{sxtn}) 
produce a clear picture of the instability
with no ``out-of-phase'' terms.  
Consider an outward radial displacement and its associated
radial magnetic field.
If $kV_A$ is not too large, the radial field is significantly 
sheared by the differential rotation to produce a negative azimuthal
elongation of the field line following equation
(\ref{sxtn}). This results in a 
positive azimuthal 
magnetic tension force, which must be balanced in the magnetostrophic
regime with an opposing Coriolis force, hence
with a yet greater radially outward displacement and magnetic field
(\ref{fftn}).   An instability is at hand.
This mechanism is illustrated in Figure~1, and 
seen through a direct simulation in Figure~2.

The growing solution for $\sigma$ is
\beq
\sigma = {|\kva|\over 2\Omega}
\left[ \left|s\, \frac{d\Omega^2}{d s} \right|
-{k^2\over k_z^2}(\kva)^2\right]^{1/2}-\eta k^2 \, .
\eeq

We wish to find the maximum growth rate of
the instability, which will be
associated with a particularly wave vector
$\bb{k}$.  The easiest way to proceed is to use the variables
\beq
X=\kva, \qquad Y={k^2\over k_z^2},
\eeq
with the understanding that $X>0$, and the parameters
\beq
\Lambda = {V_A^2\over2\eta\Omega_0} = \frac{B_0^2}{2\, \mu_0 \, \rho \, \eta\,
\Omega_0}\,, 
\quad \quad a = \left|s\, \frac{d\Omega^2}{d s}\right|
\label{eq_elsas}
\eeq
$\Lambda$ is the Elsasser number, which is typically of
order unity for the Earth's core.  
Then
\beq
2\Omega\sigma = X (a-X^2 Y)^{1/2} -X^2 Y/\Lambda \, .
\eeq
This is a decreasing function of $Y$ everywhere, with 
a maximum at the lower boundary $Y=1$.  This means that
the most rapidly growing wavenumber $k_z$ lies along
the rotation axis.  Thus
\beq
2\Omega\sigma_{\rm(max)} =  X (a-X^2 )^{1/2} -X^2 /\Lambda.
\eeq
Requiring the partial derivative $\dd/\dd X$ to vanish leads to the polynomial
\beq
X^4 -aX^2 +{a^2\over 4(1+\Lambda^{-2})}=0.
\eeq

The physically meaningful wavenumber solution is 
\beq\label{KVA}
X^2 \equiv (k_z V_A)^2 = {\Omega}\left|s\,\frac{d\Omega}{d s} \right|
\left[ 1 - (1+\Lambda^2)^{-1/2}\right] \, ,
\eeq
and corresponds to a growth rate of 
\beq\label{SIGMA}
\sigma = \left|s\, \frac{d\Omega}{d s} \right| {\Lambda/2\over
1+\sqrt{1+\Lambda^2}} \, .
\eeq

It is of interest to examine the range of $k_z$ in which instability
exists.  For the case $k=k_z$, the dispersion relation (\ref{disp})
admits growing solutions for values of $k^2_z$ less than
\beq
k^2_z = {|s\, d\Omega^2 /d s|\over V_A^2 +{4\Omega^2\eta^2/V_A^2}} \, .
\eeq
Notice that when $V_A$ is very large, $k_z$ is very small because large
$k_z$ perturbations are stabilized by magnetic tension forces.
Conversely, when $V_A$ is tiny, resistivity damps large wavenumber
perturbations.  The maximum value of $k_z$ allowing
the greatest range of unstable wavenumbers corresponds to $\Lambda =1$.
Hence, Elsasser numbers of order unity naturally emerge in an MRI-influenced
dynamo.  

At this point it is helpful to have some explicit numbers.  We take $\Omega=
7.27\times 10^{-5}$ s$^{-1}$ and $sd\Omega/ds \simeq 10^{-10}$ s$^{-1}$,
the latter corresponding to a lower bound for the angular
velocity gradient in which the characteristic
length scale associated with shear in the core is the outer core
radius itself (worse case MRI scenario).
With $\Lambda =1$ and $\eta \simeq  1$ m$^2$ s$^{-1}$, the Alfv\'en velocity is
$0.012$ m s$^{-1}$.  From equation (\ref{KVA}), we find
$$
kV_A\simeq 4.6\times 10^{-8}\ {\rm s}^{-1}
$$
which translates to a wavelength of some 1600 km.  The characteristic growth
time from equation (\ref{SIGMA}) is then about 1500 years.  
If, on the other hand, we assume that a similar jump in angular
velocity is accomodated accross a narrow shear layer (such as the
Stewartson layer, Stewartson 1966), then the characteristic growthtime
could be as short as a year.

What is the relationship between the unstable mode considered here and the  
maintenance of the
classical Taylor constraint?  It is well known that in the magnetostrophic
limit, the geostrophic component of the flow must adjust to ensure Taylor's
constraint (Taylor, 1964). A modification of the non-geostrophic flow
similar to the one envisioned here could in principle
rapidly alter the zonal shear on which the instability
itself relies. However, in the particular configuration
investigated, the wavenumber is axial.  Thus,
at least in the linear phase of the disturbance,
the Taylor constraint remains unaffected.

\section{Direct numerical simulations}

To demonstrate numerically the above mechanism, we base our study on a very
idealized model of the Earth core, chosen for illustrative purposes.
We 
consider simple spherical Couette flow driven by enforcing 
differential rotation between the inner core and the mantle.
The applied magnetic field $B_0$ is vertical and uniform.
In the simpler
hydrodynamical case, a strong shear layer will develop on the cylinder tangent 
to the inner core (Proudman, 1956; Stewartson, 1966).
With the particular choice of a vertical applied field, this shear 
is only slightly modified by MHD effects (see however Dormy {\it et al.}, 1998).
In what follows, we will work in time units of $(2\Omega_0)^{-1}$
and space units of $r_o$, the radius of the outer sphere.  It is
also convenient to introduce the dimensionless velocity
\beq
\bb{U_0} ={ \bb{V_0}\over 2r_o\Omega_0} \, ,
\eeq
where $\bb{V_0}$ is the unperturbed velocity measured in the frame
rotating at $\Omega_0$.  As before, the linear perturbed velocity
is $\bb{v}$.  For the perturbed velocity $\bb{v}$
and magnetic field $\bb{b}$, we 
introduce dimensionless variables $\bb{v'}$ and $\bb{b'}$:
\beq 
\bb{v'} = {\bb{v}/(2r_o\Omega_0)}\, , \qquad
\bb{b'} = {\bb{b}/B_0}\, .
\eeq

The governing fluid equations may then be written
  \begin{equation}
  \frac{\partial \vvv}{\partial t}+(\bb{U_0}\cdot\bfnabla)\vvv+(\vvv\cdot\bfnabla)\bb{U_0}=-\bfnabla \pi 
+ E\Delta \vvv+\frac{E \Lambda}{
  Pm}(\bfnabla\times\bb{b'})\times\bb{e_z}-\bb{e}_z\times\vvv \, ,
  \end{equation}
  \begin{equation}
  \frac{\partial \bb{b'}}{\partial
  t}=\bfnabla\times(\vvv\times\bb{e_z}+\bb{U_0}\times\bb{b'})+\
  \frac{E}{Pm}\Delta\bb{b'}
  \, ,
  \end{equation}
  \begin{equation}
  \bfnabla \cdot\vvv=\bfnabla \cdot\bb{b'}=0 \, .
  \end{equation}
In addition to the Elsasser number $\Lambda$, we have
introduced the standard Ekman and magnetic Prandtl numbers,
which are respectively:
\beq\label{EKMP}
E={\nu}/{(2\Omega_0 r_o^2)}\qquad 
Pm={\nu}/{\eta} ,
\eeq
and $\pi$ is the perturbed value of the dimesionless pressure.

As a model for the Earth's core, we investigate a spherical shell.
The angular velocity $\Omega_0$ is defined by the outer sphere.
In this reference frame, the inner sphere rotates
with angular velocity $\Omega_i>0$, in order for the 
differential rotation to decrease outward. We define the Rossby
number as 
\beq\label{RO}
Ro=\Omega_i/{\Omega_0}. 
\eeq

Let us now compare the local
WKB analysis (\ref{complete_rel_disp}) with the global 
growth rate obtained numerically.
We first need to
compute the dimensionless
steady velocity profile $\bb{U_0}$. If the magnetic field
is weak enough, this can be obtained as a solution of the hydrodynamic
problem.
The results are displayed in Table~1.
If the Elsasser number (\ref{eq_elsas}) becomes large, the field 
obviously affects the steady solution, and we then perform simulations 
with the steady MHD state as initial configuration, see Table~2.

To best illustrate the above theoretical analysis, we use numerical values as 
close as feasible to the regime described. Simulations use
$Pm=0.5$, the differential rotation is weak compared to the planet 
rotation ($Ro<\!<1$), and the Ekman number is decreased to small
values.  To accommodate small Ekman numbers,
we have implemented an axisymetric version of the Parody
code (Dormy {\it et al.}, 1998, Christensen {\it et al.}, 2001, and 
later collaborative developments) with radial resolution ranging from
$500$ to $1000$ points in radius and between  $200$ and $500$ harmonics.  

\begin{table}
\ \vskip -4.3cm 
\caption{Results obtained for the most unstable mode in a direct
integration of the governing equations. We list the growth rate
($\sigma_{\rm num}$) and the wavenumber ($k_s$ and $k_z$). We also
list the growth rate ($\sigma_{\rm th}$) obtained directly from the local
dispersion relation (\ref{complete_rel_disp}) with these wavenumber
values.  The variables $E$, $Ro$, and $\Lambda$
are defined in equations (\ref{eq_elsas}), (\ref{EKMP}) and
(\ref{RO}).
The numerical agreement is excellent.}
\begin{flushleft}
 \begin{tabular}{lcccccc} 
\tableline
 $ 2\, E$ & $Ro $ & $2\, \Lambda $ & $\sigma_{\rm th}$ & $ \sigma_{\rm num} $ & $k_s$ & $k_z$ \\
\tableline
$2\times10^{-6}$   & 0.005  & 1.0 & $6\times10^{-4}$   & $6.4\times10^{-4}$  & $\pi/0.17 $ & $ \pi/0.25$ \\
                   &        & 25  & $1.0\times10^{-3}$ & $1.09\times10^{-3}$ & $\pi/0.17$ & $\pi/0.6$ \\
$10^{-6}$          & 0.01   & 5   & $8.8\times10^{-3}$ & $1.16\times10^{-2}$ & $\pi/0.14$ & $\pi/0.15$ \\
                   & 0.01   & 25  & $1.0\times10^{-2}$ & $1.0\times10^{-2}$ & $\pi/0.14$ & $\pi/0.33$ \\
                   & 0.005  & 25  & $3.6\times10^{-3}$ & $3.6\times10^{-3}$ & $\pi/0.14$ & $\pi/0.4$ \\
                   & 0.005  & 5   & $4.2\times10^{-3}$ & $4.8\times10^{-3}$ & $\pi/0.14$ & $\pi/0.1$ \\
                   & 0.0025 & 25  & $2.0\times10^{-4}$ & $2.3\times10^{-4}$ & $\pi/0.14$ & $\pi/0.6$ \\
$5\times10^{-7}$   & 0.005  & 50  & $4.0\times10^{-3}$ & $4.4\times10^{-3}$ & $\pi/0.11$ & $\pi/0.25$ \\
                   & 0.005  & 25  & $6.0\times10^{-3}$ & $6.4\times10^{-3}$ & $\pi/0.11$ & $\pi/0.14$ \\
                   &        & 15  & $7.0\times10^{-3}$ & $6.2\times10^{-3}$ & $\pi/0.11$ & $\pi/0.125$ \\
                   &        & 10  & $7.1\times10^{-3}$ & $7.6\times10^{-3}$ & $\pi/0.11$ & $\pi/0.1$ \\
                   &        & 5   & $6.0\times10^{-3}$ & $6.4\times10^{-3}$ & $\pi/0.1$ & $\pi/0.071$ \\
                   &        & 2   & $3.0\times10^{-3}$ & $4.8\times10^{-3}$ & $\pi/0.1$ & $\pi/0.067$ \\
                   &        & 1   & $2.24\times10^{-3}$& $3.0\times10^{-3}$ & $\pi/0.1$ & $\pi/0.058$ \\
                   &        & 0.5 & $1.16\times10^{-3}$& $1.36\times10^{-3}$ & $\pi/0.11$ & $\pi/0.11$ \\
                   & 0.005  & 0.3 & $6.96\times10^{-4}$& $2.03\times10^{-4}$ & $\pi/0.1$ & $\pi/0.14$ \\
                   & 0.0025 & 25  & $2.0\times10^{-3}$ & $1.93\times10^{-3}$ & $\pi/0.1$ & $\pi/0.25$ \\
                   & 0.0075 & 25  & $6.2\times10^{-3}$ & $6.6\times10^{-3}$ & $\pi/0.1$ & $\pi/0.2$ \\
$2.5\times10^{-7}$ & 0.0025 & 25  & $2.0\times10^{-3}$ & $3.2\times10^{-3}$ & $\pi/0.08$ & $ \pi/0.15$ \\
$10^{-7}$          & 0.0025 & 1   & $3.0\times10^{-4}$ & $2.2\times10^{-4}$  & $\pi/0.06$ & $\pi/0.04$ \\
                   &        & 25  & $4.0\times10^{-3}$ & $4.8\times10^{-3}$  & $\pi/0.06$ & $\pi/0.083$ \\
\tableline
\end{tabular}
\end{flushleft}
\end{table} 

Our results are summarized in Table~1.
In agreement with theoretical expectations, the
radial wavevector 
$k_s$ varies approximately in proportion to
the dominant shear, i.e.\ $E^{1/4}$.   The
value of $k_s$ is taken from numerical simulations
by computing the width of the unstable mode at
mid-intensity, and the value of $k_z$ is obtained
by direct fourier transform.
It is found that the most unstable mode's half wavelength
generally occupies the full radial extent of the sheared region.
The inner core and the associated equatorial singularity of the Ekman
layer are the natural modal boundaries;
the equatorial singularity divides the tangent cylinder into
independent domains. 
Finally, to limit 
the stabilizing effect of diffusion, we must use $E$ much less than unity.

The induced azimuthal magnetic field clearly reproduces the
mechanism of the magnetostrophic MRI (see figure~1), despite
the presence of finite inertial and diffusive effects
(which were neglected in the theoretical calculation),
and the complications associated
with a bounded spherical domain (also neglected).

Some of these cases involve large values of the Elsasser number,
one may then worry that they are not fully self consistent. To check
this, 
we show, in Table~2, the instability parameters of such
strong field profiles, which now also have a (weak) dependence on $\Lambda$.
For the field geometry used here,
however, both the rotation profile and the applied field require relatively
little adjustment.  Agreement with the local description is also
obtained for such configurations. 

\begin{table}
\caption{Results obtained using the steady MHD state as initial
  configuration. Computations are performed with $Ro=5\times 10^{-3}$,
$Pm=0.5$ and $E=2.5 \times 10^{-7}$. As for Table~1, both the numerical 
parameters of the most unstable mode and the corresponding theoretical
  growth rate are reported.}

\begin{flushleft}
 \begin{tabular}{lcccc} 
\tableline
  $2\, \Lambda $ & $\sigma_{\rm th}$ & $ \sigma_{\rm num} $ & $k_r$ & $k_z$ \\
\tableline
3   & $5.06\times10^{-3}$ & $5.70\times10^{-3}$ & $\pi/0.92$ & $\pi/0.056$  \\
1   & $2.06\times10^{-3}$ & $1.84\times10^{-3}$ & $\pi/0.95$ & $\pi/0.056$  \\
0.3 & $8.44\times10^{-4}$ & $2.76\times10^{-4}$ & $\pi/0.10$ & $\pi/0.11$ \\
\tableline
\end{tabular}
\end{flushleft}
\end{table}

\begin{figure*}
\noindent\includegraphics[width=27pc]{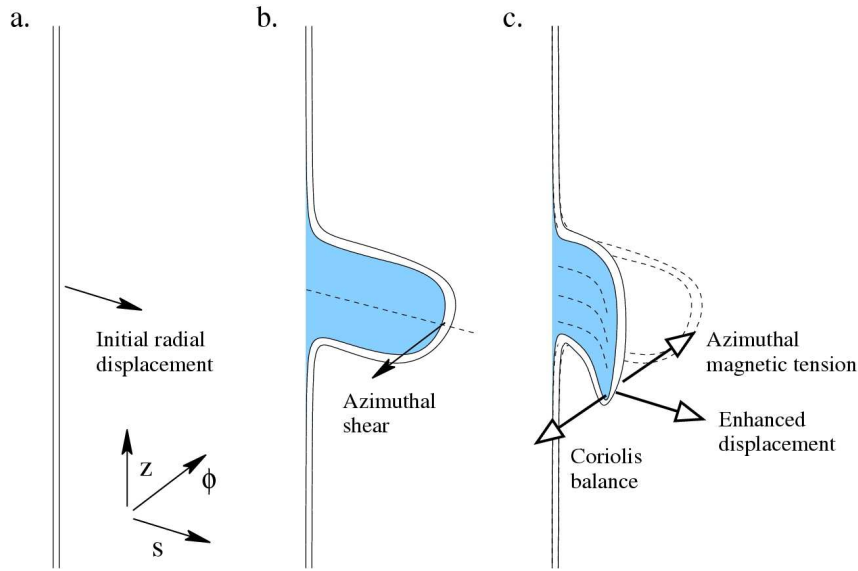}
\caption{Development of the magnetostrophic MRI.  (a)
Unperturbed field line.  (b) Field line is distorted by a radially
outward displacement, and subject to velocity shear.  (c) Field line
develops azimuthal tension which is immediately compensated by the
Coriolis force.  This compensating force requires a further displacement
in the same sense of the initial outward radial extension, and the
instability proceeds.}
\end{figure*}

\begin{figure*}
\noindent\includegraphics[width=39pc]{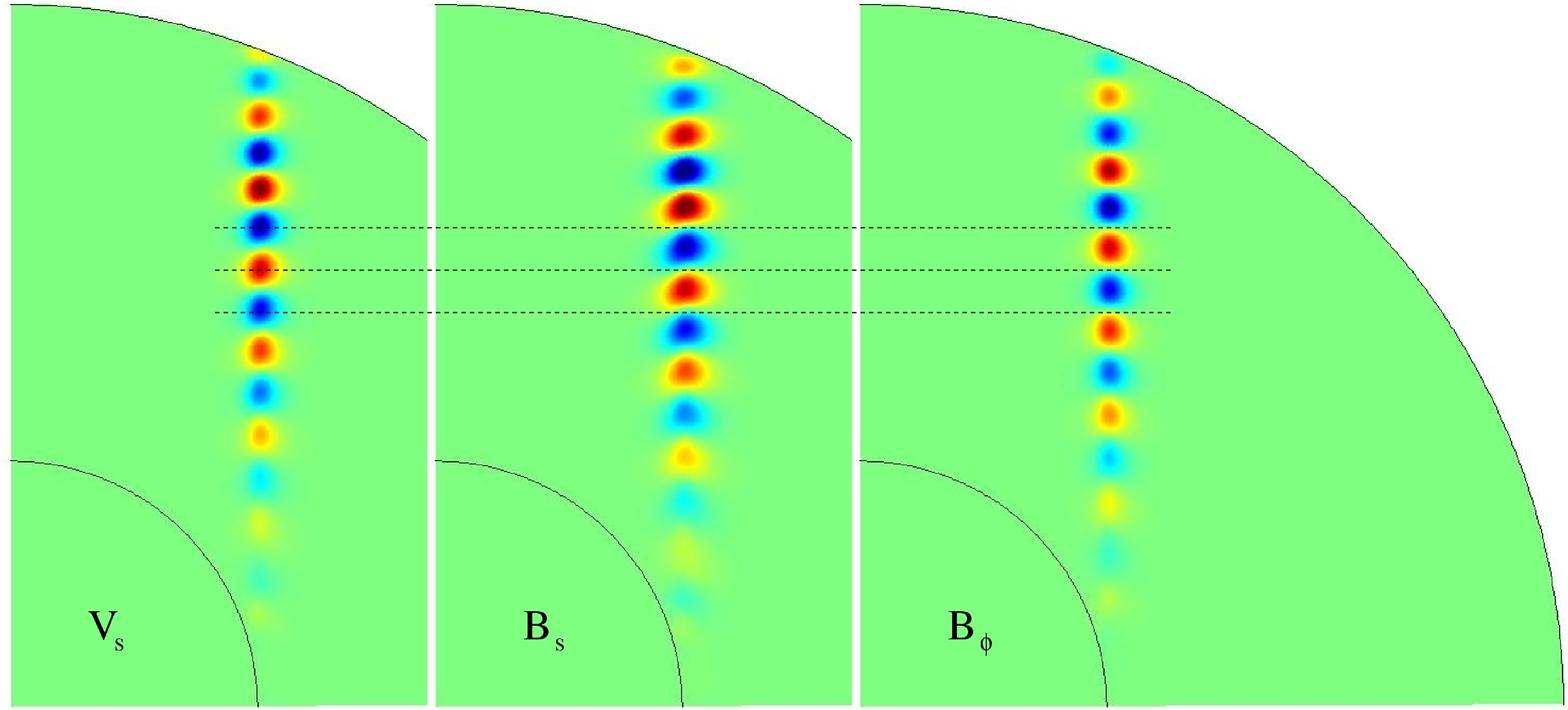}
\caption{Snapshot of the most unstable mode in a direct simulation
with $E=5 \times 10^{-7}$, $\Lambda=3$ and $Pm=0.5$. The physical
mechanism of the magnetostrophic--MRI can be traced in the phase shifts
between $v_s$, $b_s$ and $b_\phi$: $b_s$ is a quarter-period
ahead of $v_s$, while $b_\phi$ is exactly out of phase by half a
 period with $b_s$.} 
\end{figure*}

\section{Conclusions and Discussion}

At the very least, the magnetostrophic MRI discussed here seems
to be an efficient regulatory mechanism in the outer core, providing
a back reaction against the build up of angular velocity at smaller
radii.  This could be tested in full scale numerical simulations
of spherical shells whose gross aspect and parameters could
be regulated either to admit the MRI or not.  

More interestingly, might there be traces of the
magnetostrophic MRI in geomagnetic 
observations?  Because of the severe limitations in our knowledge of
differential rotation within the Earth's core, it is not possible to 
offer any kind of definitive treatment.  It is, however,
interesting to note that some axial variation in the zonal flow 
is required to account for changes in the length of the a day over 
millenial timescales (Dumberry \& Bloxham 2006). In that
respect It is interesting  that axial wavenumber
disturbances grow particularly rapidly in the model studied here.

Another temptative application is based on the observation
that the growth rate in question is directly proportional to the shear,  
which could be highly localized.  Thus,
growth times considerably shorter than 1500 years are possible:
in principle, shear times of order years could be produced. 
This suggests a
possible connection with the observations of rapid events 
of internal origin, known as ``geomagnetic jerks'' or ``geomagnetic
impulses''.  These rapid changes in the secular variation were detected and
characterized from observatory data (Courtillot {\it et al.}, 1978; Le
Mou\"el {\it et al.}, 1982) and found to correspond to localised patches of
rapid field variation at the surface of the core, using global field models
(Dormy and Mandea, 2005).  We reemphasize, however,
that the establishment of a compelling link between MRI unstable
configurations in the Earth core and geomagnetic impulses will
require considerably more effort. 
Nevertheless, the MRI offers for consideration
a novel magnetohydrodynamic mechanism that is able to induce
significant field variations over a range of observationally
interesting timescales.

%
%

\begin{acknowledgments}
We thank M. Dumberry and an anonymous referee for constructive criticisms
that greatly improved the presentation of the paper.  
Computations reported here were carried out on the CEMAG computing center
at LRA/ENS.  SAB acknowledges support from
the French Ministry of Higher Education in the form
of a Chaire d'Excellence award with additional funds from the
R\'egion Ile de France.  
\end{acknowledgments}

%
%

\end{article}


\begin{thebibliography}{}

\bibitem[]{} Acheson, D.J. (1983), Local analysis of thermal and 
magnetic instabilities in a rapidly rotating fluid, {\it 
Geophys.\ Astrophys.\ Fluid Dynam.,} {\bf 27}, 123--136

\bibitem[]{} Amit, H. and P. Olson (2004),
Helical core flow from geomagnetic secular variation,
{\it Phys. Earth Plan. Int.}, {\bf 147}, 1--25.

\bibitem[]{} Aubert, J. (2005), Steady zonal flows 
in spherical shell dynamos,
{\it J. Fluid Mech.}, {\bf 542}, 53--67.

\bibitem[]{} Balbus, S.A. and J.F. Hawley (1991) 
A powerful local shear instability in weakly magnetized disks,
{\it Astrophys. J.}, {\bf 376}, 214--222.

\bibitem[]{} 
Bloxham, J., and A. Jackson (1991),
Fluid flow near the surface of earth's outer core,
{\it Rev. Geophys.}, {\bf 29}, 97-120.

\bibitem[]{} Christensen, U. {\it et al} (2001)
A numerical dynamo benchmark,
{\it Phys. Earth Plan. Int.}, {\bf 128}, 25--34.

\bibitem[]{} Christensen, U. (2002),
Zonal flow driven by strongly supercritical convection in
rotating spherical shells, {\it J. Fluid Mech.}, {\bf 470}, 115--133.

\bibitem[]{} Courtillot, V., J. Ducruix, J.L. Le Mou\"el (1978), 
Sur une acc\'el\'eration r\'ecente de la variation s\'eculaire du champ
magn\'etique terrestre, 
{\it Acad. Sci., Paris, C.R.}, {\bf D287}, 1095-1098.

\bibitem[]{} Dormy, E., P. Cardin and D. Jault (1998),
MHD flow in a slightly differentially rotating
spherical shell, with conducting inner core,
in a dipolar magnetic field,
{\it Earth and Planetary Science Letters}, {\bf 160},
15--30.

\bibitem[]{} Dumberry, M. (2007), Geodynamics constraints
on the steady  and time-dependent inner core axial rotation, 
{\it Geophys.\ J.\ Int.}, {\bf 170}, 886--895.                 

\bibitem[]{} Dumberry, M. and J. Bloxham (2006), Azimuthal
flows in the Earth's core and changes in the length of the day at 
millenial time scales, 
{\it Geophys.\ J.\ Int.}, {\bf 165}, 32--46.                  

\bibitem[]{} Dormy, E. and M. Mandea (2005),
Tracking geomagnetic impulses at the core-mantle boundary,
{\it Earth Planet. Sci. Lett.}, {\bf 237},  300--309.

\bibitem[]{} Eymin, C. and G. Hulot (2005),
On core surface flows inferred from satellite magnetic data,
{\it Phys. Earth Plan. Int.}, {\bf 152}, 200--220.

\bibitem[]{} Fearn, D.R., C.J. Lamb, D.R. McLean,R.R. Ogden (1997),
The influence of differential rotation on magnetic instability,
and nonlinear magnetic instability in the magnetostrophic limit,
{\it Geophys.\ Astrophys.\ Fluid Dynam.}, {\bf 86}, 173--200.

\bibitem[]{} Le Mou\"el, J.L., J. Ducruix, C. Ha Duyen (1982), 
The worldwide character
of the 1969-1970 impulse of the secular acceleration rate, 
{\it Phys. Earth Planet. Inter.}, {\bf 28}, 337-350.

\bibitem[]{} Menou K., S.A. Balbus, and H.C. Spruit (2004),
Local axisymmetric diffusive stability of weakly magnetized,
differentially rotating, stratified fluids,
{\it Astrophys. J.}, {\bf 607}, 564--574.

\bibitem[]{}
Ogden, R.R., D.R. Fearn (1995), The destabilising role of differential 
rotation, {\it Geophys.\ Astrophys.\ Fluid Dynam.}, {\bf 81}, 215--232.

\bibitem[]{} Proudman, I. (1956), The almost-rigid rotation of viscous fluid
between concentric spheres, {\it J. Fluid Mech.}, {\bf 1}, 505--516.

\bibitem[]{} Stewartson, K. (1966), On almost rigid rotations, Part 2, {\it
  J. Fluid Mech.}, {\bf 26}, 131--144.

\bibitem[]{} Taylor, J.B. (1964), The Magneto-Hydrodynamics of a Rotating Fluid
  and the Earth's Dynamo Problem, {\it Proc. Roy. Soc. Lond.}, {\bf 274},
  274--283. 

\bibitem[]{} Vidale, J. E., D. A. Dodge, and P.S. Earle (2000), 
Slow differential rotation of the Earth's inner core indicated by temporal
changes in scattering, {\it Nature}, {\bf 405}, 445--448. 




\end{thebibliography}
\end{document}